\documentclass[%
 reprint,
superscriptaddress,
 amsmath,amssymb,
 aps,
]{revtex4-1}

\usepackage{graphicx}% Include figure files
\usepackage{dcolumn}% Align table columns on decimal point
\usepackage{bm}% bold math
\usepackage{subfigure}
\begin{document}

\preprint{APS/123-QED}

\title{Continuous-variable measurement-device-independent
quantum key distribution via quantum catalysis}% Force line breaks with \\

\author{Wei Ye}%
\affiliation{School of Computer Science and Engineering, Central South University, Changsha 410083, China}

\author{Hai Zhong}%
\affiliation{School of Computer Science and Engineering, Central South University, Changsha 410083, China}

\author{Xiaodong Wu}%
\affiliation{School of Computer Science and Engineering, Central South University, Changsha 410083, China}

\author{Liyun Hu}%
\email{Corresponding author:  hlyun2008@126.com}\affiliation{Center for Quantum Science and Technology, Jiangxi Normal University, Nanchang 330022, China}

\author{Ying Guo}%
\email{Corresponding author:  yingguo@csu.edu.cn}\affiliation{School of Computer Science and Engineering, Central South University, Changsha 410083, China}

\date{\today}% It is always \today, today,
             %  but any date may be explicitly specified

\begin{abstract}
The continuous-variable measurement-device-independent quantum key
distribution (CV-MDI-QKD) is a promising candidate for the immunity to
side-channel attacks, but unfortunately seems to face the limitation of
transmission distance in contrast to discrete-variable (DV)
counterpart. In this paper, we suggest a method of improving the performance
of CV-MDI-QKD involving the achievable secret key rate and transmission
distance by using zero-photon catalysis (ZPC), which is indeed a
noiseless attenuation process. The numerical stimulation results show that
the transmission distance of ZPC-based CV-MDI-QKD under the extreme
asymmetric case is better than that of the original protocol. Attractively,
in contrast to the previous single-photon subtraction (SPS)-based
CV-MDI-QKD, the proposed scheme enables a higher secret key
rate and a longer transmission distance. In particular, the ZPC-based
CV-MDI-QKD can tolerate more imperfections of detectors than both
the original protocol and the SPS-based CV-MDI-QKD.
\end{abstract}

\maketitle

\section{Introduction}

Quantum key distribution (QKD) \cite{1,2,3,4} is one of the most mature
domains of quantum information processing, aiming to establish a shared key
between two distance honest parties (Alice and Bob) over an insecure channel
controlled by an eavesdropper (Eve), and its security can be ensured by
quantum mechanics. A review of the latest developments in quantum
cryptography can be found in \cite{5}. In particular, the first theoretical
Bennett-Brassard 1984 (BB84) protocol \cite{6} was proposed, so that the
discrete-variable (DV) QKD \cite{6,7,8} has received increasing attention,
and are even available on the commerce. It can perform outstandingly with
respect to the transmission distance, but may suffer from the restriction of
lower secret key rates due to the dependence of the single-photon generation
and detection.

In order to overcome this shortcoming, the continuous-variable (CV) QKD \cite%
{9,10,11,12,13} has emerged as a new solution to promise higher secret key
rates with the help of the homodyne or heterodyne detection rather than
photon counters, which make it more attractive from a practical viewpoint.
Especially, the coherent-state Gaussian modulated CVQKD \cite{9} has been
rigorously proved to be secure against arbitrary collective attacks \cite{14}%
, which are optimal in the asymptotical limit \cite{15}. Moreover, it has an
advantage of compatibility with traditional telecommunication technologies,
and thus shows the potential to be used for the next-generation quantum
communication networks \cite{16}. Unfortunately, when considering the
imperfection of the detector from a realistic scenario, it opens the door to
potential security loopholes that could be successfully exploited by Eve to
execute attack strategies, such as the local oscillator calibration attack
\cite{17}, the wavelength attack \cite{18}, and the detector saturation
attack \cite{19}. To resist these attacks, there are usually two solutions,
i.e., the device-independent QKD \cite{20,21} and the
measurement-device-independent (MDI) QKD \cite{22,23,24,25,26,27}. Different
from the former that based on the violation of a Bell inequality \cite{20},
the latter is a more practical way to prevent all side-channel attacks on
detection where the security of the protocol does not rely on the
reliability of the measurement devices. Even so, when comparing with that of
DV-MDI-QKD \cite{28,29}, the maximal transmission distance of CV-MDI-QKD is
still unsatisfactory. Thus, how to effectively improve the maximal
transmission distance in CV-MDI-QKD is an interesting and challenging task.

Till now, many efforts have been devoted to improving its performance in
CV-MDI-QKD systems. In general, the use of discrete modulations \cite{30} or
quantum operations \cite{31,32,33}\ is a viable means. For instance, a
discrete-modulated CV-MDI-QKD protocol has been proposed recently, which
outperforms the Gaussian-modulated CV-MDI-QKD protocol with respect to the
achievable maximal transmission distance since such a discrete modulation
has efficient reconciliation error correction codes in the regime of low
signal-to-noise ratio \cite{30}. However, it has a problem that the
modulation variance should be sufficient low in order to derive the Eve's
Holevo information, which may cause the transmitting power of the quantum
signal to be too low and hence has much effect on the performance of QKD. In
addition, a novel CV-MDI-QKD protocol using optical amplifiers has shown
that it can achieve a higher secret key rate and a longer transmission
distance, compared with previous coherent- and squeezed-states protocols
\cite{31}. Recently, the photon-subtraction operation \cite{34,35}, which
can be emulated by non-Gaussian postselection \cite{36}, has been proved to
lengthen the maximal transmission distance of CV-MDI-QKD \cite{32,33}. More
interesting, the single-photon subtraction (SPS) presents the best
performance. In spite of the photon subtraction showing its unique
advantages, however, there are still restricted to the low success
probability of implementing such an operation for a given certain variance
of the Einstein-Podolsky-Rosen (EPR) state, thereby resulting in the loss of
information between Alice and Bob during the distillation of secret keys.
Fortunately, the quantum catalysis operation \cite{37}, which can be
implemented with existing technologies, may become an alternative method of
improving the performance of CVQKD systems in terms of secret key rate and
transmission distance, especially in the case of zero-photon catalysis (ZPC)
\cite{38,39}. Currently, the characteristics of quantum catalysis have been
widely utilized in quantum coherence \cite{40}, nonclassicality \cite{41},
entanglement property \cite{42,43}, and so on. As far as we know, there is
few applications of quantum catalysis in CV-MDI-QKD. Inspired by the
afore-mentioned analysis, in this paper, we suggest a method to improve the
performance of coherent-state CV-MDI-QKD by using the ZPC, which has the
characteristics of noiseless attenuation and can keep the Gaussian behavior
of Wigner function. The proposed ZPC-based CV-MDI-QKD can lengthen the
maximal transmission distance with the achievable high secret key rate. It
performs better than the SPS case with respect to both secret key rate and
transmission distance.

This paper is structured as follows. In Sec. II, we describes the
characteristics of the CV-MDI-QKD protocol involving the ZPC operation. In
Sec.III, we show the performance of the ZPC-based CV-MDI-QKD system. The
secret key rate of the ZPC-based CV-MDI-QKD is first derived according to
the optimality of Gaussian attack. After that, the simulations and
performance analysis results are provided. Finally, conclusions are drawn in
Sec.IV.

\section{The ZPC-based CV-MDI-QKD protocol}

\begin{figure}[ptb]
\centering \includegraphics[width=\columnwidth]{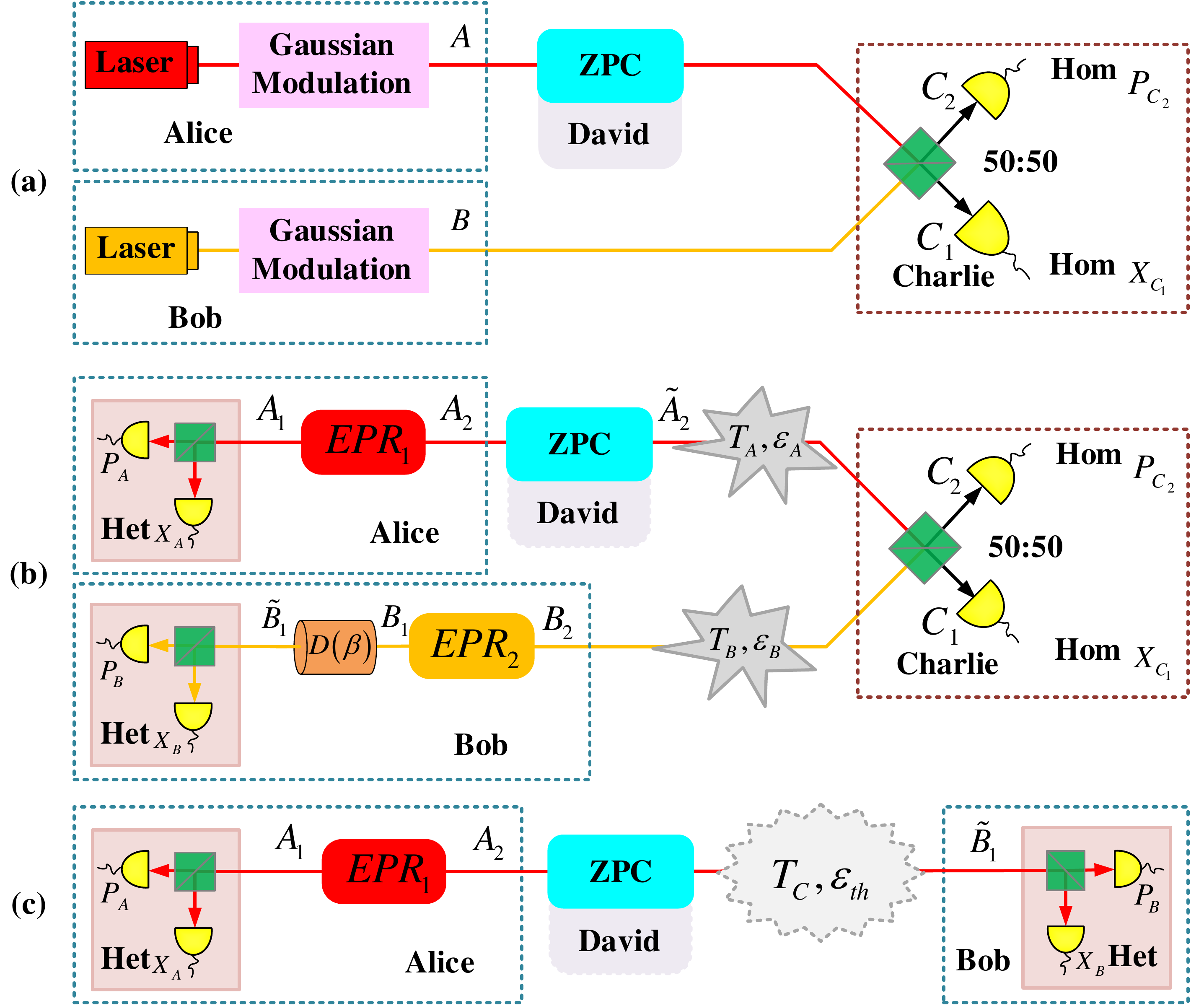}
\caption{ (Color
online) Schematic diagram of the CV-MDI-QKD protocol with the ZPC operation (cyan box).
(a) Prepare-and-measure (PM) scheme of the ZPC-based
CV-MDI-QKD. (b) Entanglement-based (EB) scheme of the ZPC-based CV-MDI-QKD.
(c) Equivalent one-way protocol of the EB scheme of the ZPC-based CV-MDI-QKD
under the assumption that Eve is aware of David, Charlie, and Bob's EPR$_{2}$
state and displacement except for heterodyne detection. EPR$_{1}$ and
EPR$_{2}$: Alice's and Bob's two-mode squeezed state, respectively. Het: heterodyne detection. Hom: homodyne detection. $\left \{X_{A},P_{A}\right \}  $ and $\left \{  X_{B},P_{B}\right \}  $: Alice's and Bob's
measurement results of heterodyne detection, respectively. $X_{C_{1}}%
,P_{C_{2}}$: measurement results of homodyne detection of measuring the $X$
and $P$ quadrature, respectively. $\widehat{\Pi}^{off}$: projection operator
$\left \vert 0\right \rangle \left \langle 0\right \vert $. $T_{A}(\varepsilon
_{A})$, $T_{B}\left(  \varepsilon_{B}\right)  $: channel parameters for
Alice-Charlie and Bob-Charlie. $T_{c}(\varepsilon_{th})$: equivalent channel
transmittance (excess noise). $D\left(  \beta \right)  $: displacement
operation of Bob.  }
\label{Fig1}
\end{figure}

In order to improve the performance of the CVQKD system, we elaborate the ZPC-based CV-MDI-QKD protocol with
Gaussian-modulated coherent states in Fig. 1. Among them, Fig. 1(a) shows the
prepare-and-measure (PM) scheme of the ZPC-based CV-MDI-QKD protocol. It is easy to implement the PM scheme in practice, but not conducive to security
analysis. Consequently, we consider its equivalent
entanglement-based (EB) scheme of the ZPC-based CV-MDI-QKD, as depicted in
Fig. 1(b), where Alice and Bob respectively prepare an entanglement resource,
i.e., EPR$_{1}$ and EPR$_{2}$ with variances $V_{A}$ and $V_{B}$. They
retain modes $A_{1}$ and $B_{1}$, and send other modes $A_{2}$ and $%
B_{2}$ to an untrusted third party Charlie through the quantum channel with
length $L_{AC}$ and $L_{BC}$, respectively. To reduce equipment
requirements, we assume that the ZPC operation is controlled by an
untrusted party David, who is close to Alice's station. On top of that,
Charlie, first interferes with two received modes $\widetilde{A}_{2}$ and $%
B_{2}$ via a $50$:$50$ beam splitter and obtains two output modes $C_{1}$
and $C_{2}$, and then performs homodyne detection to obtain a measurement
results $\left\{ X_{C_{1}},P_{C_{2}}\right\} $, which are publicly announced
through a classical channel. After receiving $\left\{
X_{C_{1}},P_{C_{2}}\right\} $, Bob uses a displacement operation $D\left(
\beta \right) $ with $\beta =g\left( X_{C_{1}}+iP_{C_{2}}\right) $ to modify
mode $B_{1}$ to $\widetilde{B}_{1}$, where $g$ is a gain factor. By using
heterodyne detection, Alice and Bob respectively measure modes $A_{1}$ and $%
\widetilde{B}_{1}$ to obtain their own data $\left\{ X_{A},P_{A}\right\} $
and $\left\{ X_{B},P_{B}\right\} $ of which the data can be used for
implementing parameter estimations. Finally, a string of secret key can be
extracted via data post-processing.

\begin{figure}[t]
\centering
\subfigure[]{
\centering
\includegraphics[width=0.81\columnwidth]{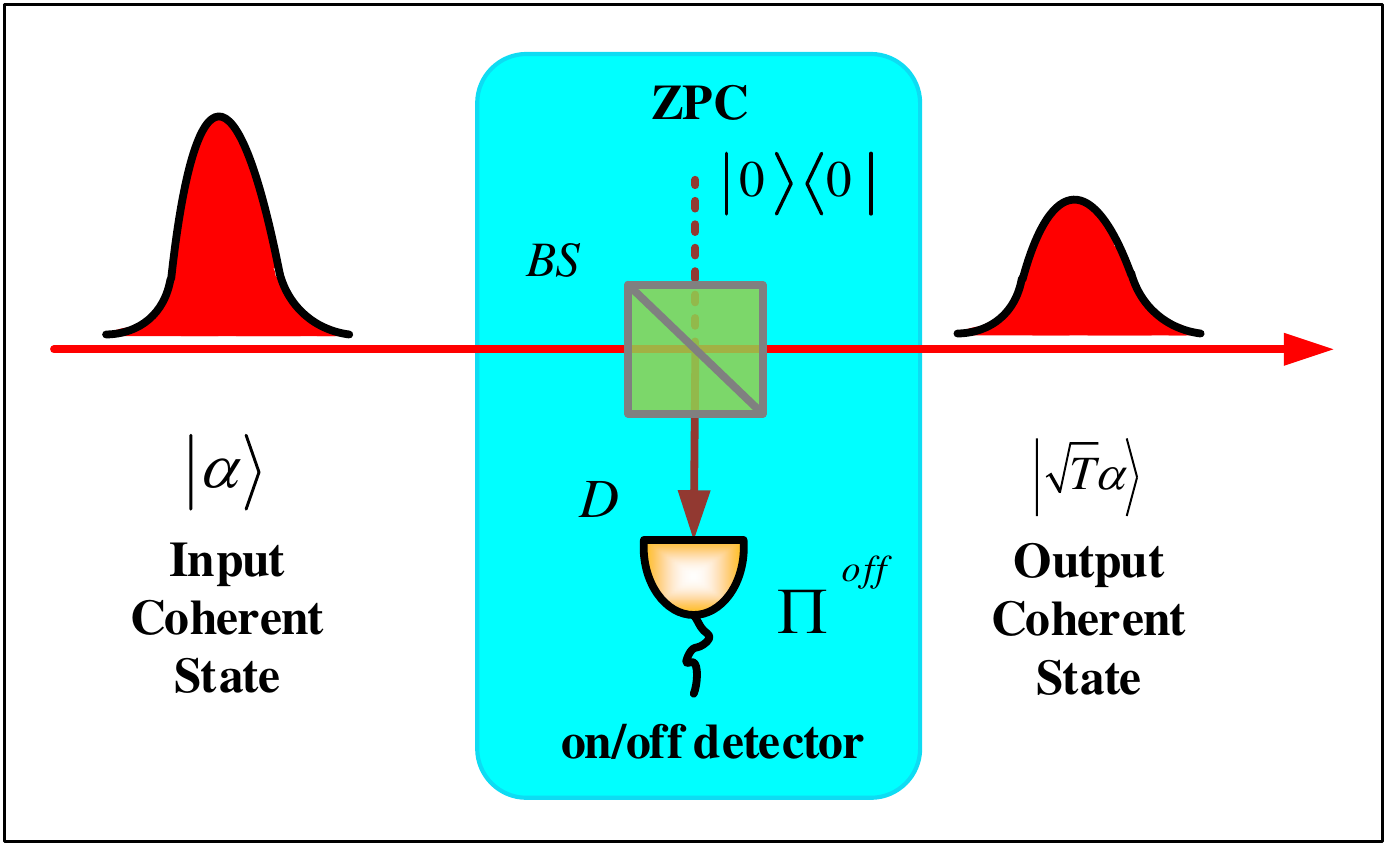}
\label{Fig2a}
}
\subfigure[]{
\includegraphics[width=0.85\columnwidth]{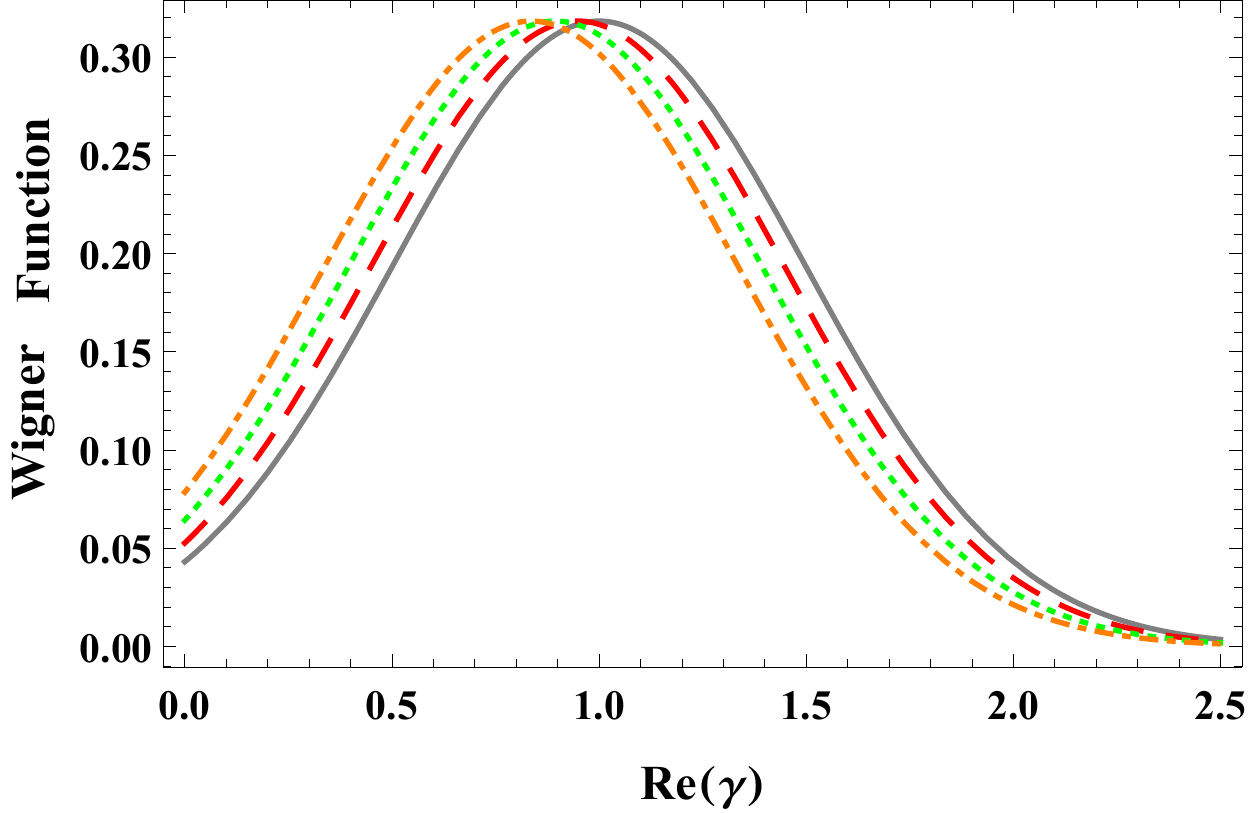}
\label{Fig2b}
}
\caption{ (Color online) (a) Schematic structure of the ZPC (cyan box) in the PM scheme of
CV-MDI-QKD.   BS: beam splitter with a transmittance $T$. $\left \vert
\alpha \right \rangle $: coherent state. $\widehat{\Pi}^{off}$: projection
operator $\left \vert 0\right \rangle \left \langle 0\right \vert $. (b)  Wigner
function of $\left \vert \sqrt{T}\alpha \right \rangle $ and $\left \vert
\alpha \right \rangle $ at a given amplitude $\left \vert \alpha \right \vert =1$ as
a function of $\operatorname{Re}\left(  \gamma \right)  $ in the phase space
$\gamma \in \left(  q,p\right)  $ with several different values of $T.$The
highest peaks from right to left correspond to $T\in \left \{1,0.9,0.8,0.7\right \},$
respectively. For convenience, here we take $p=0$, and the black line
represents the input coherent state $\left \vert \alpha \right \rangle $.}
\label{Fig2}
\end{figure}

Due to the assumption that both the EPR$_{2}$ state and the displacement
operation $D\left( \beta \right) $ are untrusted, the EB scheme of the
ZPC-based\ CV-MDI-QKD can be equivalent to that of the one-way protocol
using heterodyne detection \cite{12,24,32,33}, which is illustrated in Fig.
1(c). It is worth noting that, different from the equivalent one-way
protocol, the model of CV-MDI-QKD has two lossy quantum channels. Thus, from
the point of view of the attack strategies, Eve can take two attacks, e.g.
one-mode attack and two-mode attack. However, in a practical system,
adopting a two-mode attack is difficult to be achieved thanks to the
immaturity of quantum memory technologies. Furthermore, if two lossy quantum
channels are independent, then two-mode attack can be reduced to the
one-mode attack. As a result, one can take the security analysis under two
Markovian memoryless Gaussian channels where no entanglement can be
distributed, so that in CV-MDI-QKD, two quantum channels degenerate into the
one-mode channel \cite{44}. In this situation, the optimal attack strategy
for Eve is regarded as one-mode collective Gaussian attack.

Now, let us turn attention to the relationship of channel parameters between
the ZPC-based CV-MDI-QKD and its equivalent one-way protocol. In Fig. 1(b),
since the channels of both Alice-Charlie and Bob-Charlie are both linear,
these channels that controlled by Eve can be emulated by using two
independent entangling cloner attacks where $T_{A}=10^{-\kappa L_{AC}/10}$ ($%
T_{B}=10^{-\kappa L_{BC}/10}$) and $\varepsilon _{A}$ ($\varepsilon _{B}$)
represent the transmittance and excess noise of the channel between Alice
(Bob) and Charlie, $\kappa =0.2$ dB/km. In Fig. 1(c), $T_{c}$ and $%
\varepsilon _{th}$ respectively represent the transmittance and excess noise
of the equivalent one-way protocol, i.e., $T_{c}={g^{2}T_{A}}/{2}$, and $%
\varepsilon _{th}=\left( \sqrt{2V_{B}-2}/{g}-\sqrt{T_{B}V_{B}+T_{B}}\right)
^{2}/{T_{A}}+{T_{B}}\left( \chi _{B}-1\right) {T_{A}}+\chi _{A}+1,$ where $%
\chi _{j}=\left( 1-T_{j}\right) /T_{j}+\varepsilon _{j}$ with $j\in \{A,B\}$%
. In order to minimize the equivalent excess noise $\varepsilon _{th}$, we
take into account $g^{2}=2\left( V_{B}-1\right) /\left[ T_{B}\left(
V_{B}+1\right) \right] $, and then have excess noise given by
\begin{equation}
\varepsilon _{th}=\frac{T_{B}}{T_{A}}\left( \varepsilon _{B}-2\right)
+\varepsilon _{A}+\frac{2}{T_{A}}.  \label{1}
\end{equation}

From a practical point of view, we need to consider Charlie's inefficient
detections, which can be characterized by a quantum efficiency $\eta $ and
an electronic noise $v_{el}$. Thus, the detection-added noise can be given
by $\chi _{\hom }=\left( v_{el}+1-\eta \right) /\eta $, and the total noise
referred to the channel input is expressed as $\chi _{tot}=$ $\chi
_{line}+2\chi _{\hom }/T_{A}$ where $\chi _{line}=\left( 1-T_{c}\right)
/T_{c}+\varepsilon _{th}$ refers to the channel-added noise. Note that all
the above noises are in the shot-noise units (SNU).

Before evaluating the performance of the ZPC-based CV-MDI-QKD system, we
suggest the physical characteristics of quantum catalysis. As shown in Fig.
2, the specific structure of ZPC operation (cyan box) is that a vacuum state
$\left\vert 0\right\rangle _{D}$ in auxiliary mode $D$ is sent to a beam
splitter (BS) with a transmittance $T=1-R$, and subsequently an on/off
detector only registers zero-photon (no click). Such a catalytic process
can, in fact, be taken as an equivalent operator $\hat{O}_{0}$
\begin{equation}
\hat{O}_{0}\equiv \text{T}r\left[ B\left( T\right) \widehat{\Pi }_{off}%
\right] =\sqrt{T}^{a_{2}^{\dagger }a_{2}},  \label{2}
\end{equation}%
where $B\left( T\right) $ is the normal ordering form of a BS operator given
by $\exp [(\sqrt{T}-1)(a_{2}^{\dagger }a_{2}+d^{\dag }d)+$($d^{\dag
}a_{2}-da_{2}^{\dagger }$)$\sqrt{R}]$, and $\widehat{\Pi }_{off}=\left\vert
0\right\rangle _{D}\left\langle 0\right\vert $ is one of the projection
operators in the on/off detector.

After performing by David the ZPC operation for the incoming EPR$_{1}$ state
in mode $A_{2}$, where EPR$_{1}$ on Alice's side can be prepared by applying
a two-mode squeezed operator $S_{A_{1}A_{2}}\left( r\right) =\exp $[$r$($%
a_{1}a_{2}-a_{1}^{\dagger }a_{2}^{\dagger }$)] with a squeezing parameter $r$
into the two-mode vacuum state, i.e.,%
\begin{align}
\left\vert EPR_{1}\right\rangle _{A_{1}A_{2}}& =S_{A_{1}A_{2}}\left(
r\right) \left\vert 00\right\rangle _{A_{1}A_{2}}  \notag \\
& =\sqrt{1-\lambda ^{2}}\exp \left( \lambda a_{1}^{\dagger }a_{2}^{\dagger
}\right) \left\vert 00\right\rangle _{A_{1}A_{2}},  \label{3}
\end{align}%
where $\lambda =\sqrt{\left( V_{A}-1\right) /\left( V_{A}+1\right) }$ with $%
V_{A}=\cosh 2r$. The resulting state $\left\vert \Phi \right\rangle _{A_{1}%
\widetilde{A}_{2}}$ can be described as $\ $%
\begin{align}
\left\vert \Phi \right\rangle _{A_{1}\widetilde{A}_{2}}& =\frac{\hat{O}_{0}}{%
\sqrt{P_{d}}}\left\vert EPR_{1}\right\rangle _{A_{1}A_{2}}  \notag \\
& =\sqrt{\frac{1-\lambda ^{2}}{P_{d}}}\exp \left( \lambda \sqrt{T}%
a_{1}^{\dagger }a_{2}^{\dagger }\right) \left\vert 00\right\rangle _{A_{1}%
\widetilde{A}_{2}},  \label{4}
\end{align}%
with the normalization factor $P_{d}$%
\begin{equation}
P_{d}=\frac{2}{1+T+RV_{A}},  \label{5}
\end{equation}%
representing the success probability of implementing the ZPC. Based on Eq. (%
\ref{4}), the covariance matrix of the state $\left\vert \Phi \right\rangle
_{A_{1}\widetilde{A}_{2}}$ can be calculated as%
\begin{equation}
\Gamma _{A_{1}\widetilde{A}_{2}}=\left(
\begin{array}{cc}
x\Pi  & z\sigma _{z} \\
z\sigma _{z} & y\Pi
\end{array}%
\right) ,  \label{6}
\end{equation}%
where $\Pi $ represents two-dimensional identity matrix, $\sigma_{z}=diag\left( 1,-1\right) $, and $x$, $y$, $z$ are given by%
\begin{align}
x& =y=\frac{2V_{A}-RV_{A}+R}{1+T+RV_{A}},  \notag \\
z& =\frac{2\sqrt{T\left( V_{A}^{2}-1\right) }}{1+T+RV_{A}}.  \label{7}
\end{align}

\begin{figure}[ptb]
\label{Fig3} \centering \includegraphics[width=0.8\columnwidth]{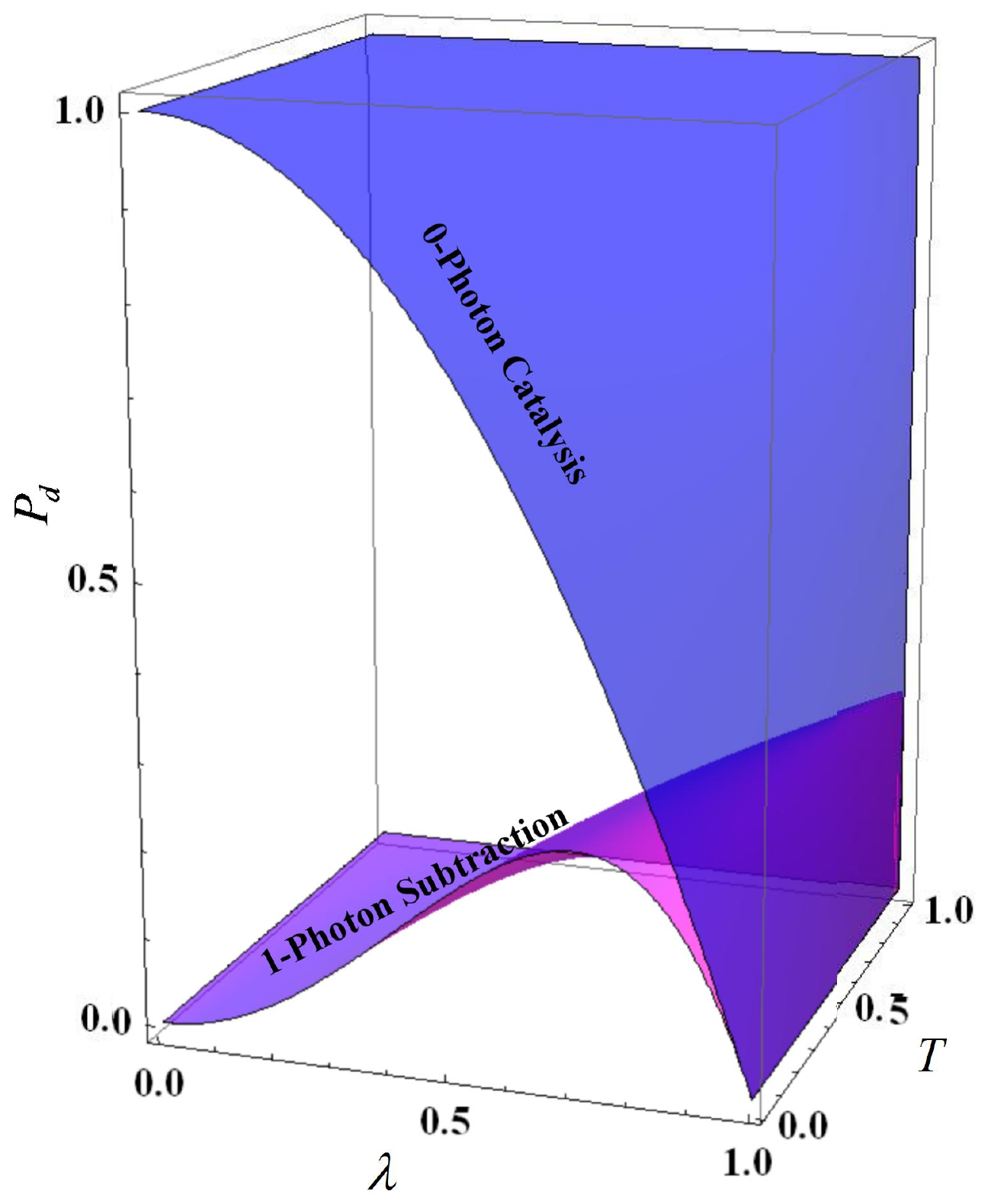}
\caption{(Color online) The success probability $P_{d}$ of the ZPC operation as a function of
$T$ and $\lambda$. As a comparison, the magenta surface represents the SPS
case.}
\end{figure}

As shown in Eq. (\ref{2}), the ZPC happens to be a noiseless attenuation
described as $\hat{O}_{0}\left\vert \alpha \right\rangle \rightarrow
\left\vert \sqrt{T}\alpha \right\rangle $, where $\left\vert \alpha
\right\rangle $ is the coherent state in the PM scheme of CV-MDI-QKD.To this
end, we plot the Wigner function between the input and output states at a
fixed amplitude $\left\vert \alpha \right\vert =1$ as a function of $\operatorname{Re}\left(  \gamma \right)  $ in the phase space $\gamma
\in \left(  q,p\right)  $ 
with several different values of $T\in \left\{ 1,0.9,0.8,0.7\right\} ,$ as
shown in Fig. 2b. We can find that the ZPC not only can effectively maintain
the Gaussian behavior of Wigner function, but also does not introduce noises
in terms of the Gaussian distribution between $\left\vert \alpha
\right\rangle $ and $\left\vert \sqrt{T}\alpha \right\rangle .$ In addition,
unlike the photon-subtraction operation, the ZPC can facilitate the
transformation of the target ensemble between modes $A_{1}$ and $\widetilde{A%
}_{2}$ since the auxiliary vacuum state itself keeps unaffected in mode $D$.
To clearly see this viewpoint, Fig. 3 illustrates the success probabilities
of both ZPC (blue surfance) and SPS (magenta surface) with different
transmittances $T$ at each $\lambda $. We find that the success probability
of implementing the ZPC is always better than the SPS case and the gap at a
given transmittance $T$ extends with the decrease of $\lambda $. This means
that the ZPC has the advantage of the success probability over the
photon-subtraction case, thereby effectively preventing data loss between
Alice and Bob in the process of extracting the secret key. Moreover, after
the ZPC, interestingly, the resulting state $\left\vert \Phi \right\rangle
_{A_{1}\widetilde{A}_{2}}$ in Eq. (\ref{4}) is still an EPR state with an
updated squeezing parameter $\lambda \sqrt{T}$.

\section{Performance analysis of the ZPC-based CV-MDI-QKD}

So far we have suggested the structure characteristics of the ZPC-based CV-MDI-QKD system. In what follows, we calculate the asymptotic secret key rate under the equivalent one-way CVQKD protocol
using heterodyne detection, where Bob performs a reverse reconciliation. We analyze the security  through  numerical simulations and demonstrate the performance improvement of the ZPC-based CV-MDI-QKD system in terms of secret key rate and transmission distance.

\subsection{Derivation of the secret key rate}

It is interesting to note that, after performing the ZPC, the traveling
state $\left\vert \Phi \right\rangle _{A_{1}\widetilde{A}_{2}}$ is still a
Gaussian state, which makes it suitable to directly derive the secret key
rate from the conventional Gaussian CVQKD. According to the optimality of
Gaussian attack \cite{45,46,47}, one can calculate the asymptotic secret key
rate $K$ by using the covariance matrix in Eq. (\ref{6}). Subsequently, the
asymptotic secret key rate of the equivalent one-way ZPC-based CVQKD system
for reverse reconciliation against one-mode collective Gaussian attack is
given by
\begin{equation}
K=P_{d}\left\{ \beta I\left( A\text{:}B\right) -\chi \left( B\text{:}%
E\right) \right\} ,  \label{8}
\end{equation}%
where $P_{d}$ has been defined in Eq. (\ref{5}), $\beta $ is the
reverse-reconciliation efficiency, $I\left( A\text{:}B\right) $ denotes the
Shannon mutual information between Alice and Bob, and $\chi \left( B\text{:}%
E\right) $ represents the Holevo bound between Bob and Eve.

As shown in Fig. 1(c), when the state $\left\vert \Phi \right\rangle _{A_{1}%
\widetilde{A}_{2}}$ passes through an untrusted quantum channel
characterized by $T_{c}$ and $\varepsilon _{th}$, the covariance matrix of
the state $\left\vert \Phi \right\rangle _{A_{1}\widetilde{B}_{1}}$ can be
described as follows%
\begin{eqnarray}
\Gamma _{A_{1}\widetilde{B}_{1}} &=&\left(
\begin{array}{cc}
X\Pi  & Z\sigma _{z} \\
Z\sigma _{z} & Y\Pi
\end{array}%
\right)   \notag \\
&=&\left(
\begin{array}{cc}
x\Pi  & \sqrt{T_{c}}z\sigma _{z} \\
\sqrt{T_{c}}z\sigma _{z} & T_{c}\left( x+\chi _{tot}\right) \Pi
\end{array}%
\right) .  \label{9}
\end{eqnarray}%
Thus, the Shannon mutual information $I\left( A\text{:}B\right) $ can be
calculated as
\begin{align}
I\left( A\text{:}B\right) & =\log _{2}\frac{V_{A_{M}}}{V_{A_{M}|B_{M}}}
\notag \\
& =\log _{2}\frac{\left( X+1\right) \left( Y+1\right) }{\left( X+1\right)
\left( Y+1\right) -Z^{2}}.  \label{10}
\end{align}%
In order to obtain the Holevo bound $\chi \left( B\text{:}E\right) $, we
assume that Eve perceives the existence of the untrusted party David and
purifies the whole system $\rho _{A_{1}\widetilde{B}_{1}ED}$, so that
\begin{align}
\chi \left( B\text{:}E\right) & =S\left( E\right) -S\left( E|B\right)
\notag \\
& =S\left( A_{1}\widetilde{B}_{1}\right) -S\left( A_{1}|\widetilde{B}%
_{1}^{m_{B}}\right) ,  \notag \\
& =\underset{i=1}{\overset{2}{\sum }}G\left( \frac{\lambda _{i}-1}{2}\right)
-G\left( \frac{\lambda _{3}-1}{2}\right) ,  \label{11}
\end{align}%
with the von Neumann entropy%
\begin{equation}
G\left( \varsigma \right) =\left( \varsigma +1\right) \log _{2}\left(
\varsigma +1\right) -\varsigma \log _{2}\varsigma ,  \label{12}
\end{equation}%
where $S(A_{1}\widetilde{B}_{1})$ is a function of the symplectic
eigenvalues $\lambda _{1,2}$ of $\Gamma _{A_{1}\widetilde{B}_{1}}$ given by $%
\lambda _{1,2}^{2}=(\Delta \pm \sqrt{\Delta ^{2}-4\xi ^{2}})/2,$ with $%
\Delta =X^{2}+Y^{2}-2Z^{2}$ and $\xi =XY-Z^{2}$, and Eve's condition entropy
$S(A_{1}|\widetilde{B}_{1}^{m_{B}})$ based on Bob's measurement result $m_{B}
$, is a function of the symplectic eigenvalues $\lambda _{3}$ of $\Gamma
_{A_{1}}^{B_{1}^{\prime m_{B}}}=\Gamma _{A_{1}}-\sigma _{A_{1}B_{1}^{\prime
}}(\Gamma _{B_{1}^{\prime }}+\Pi )^{-1}\sigma _{A_{1}B_{1}^{\prime }}^{\text{%
T}}$, which is given by $\lambda _{3}=X-{Z^{2}}/({Y+1}).$

\subsection{Simulation results and analysis}

\begin{figure}[t]
\centering
\subfigure[]{
\centering
\includegraphics[width=\columnwidth]{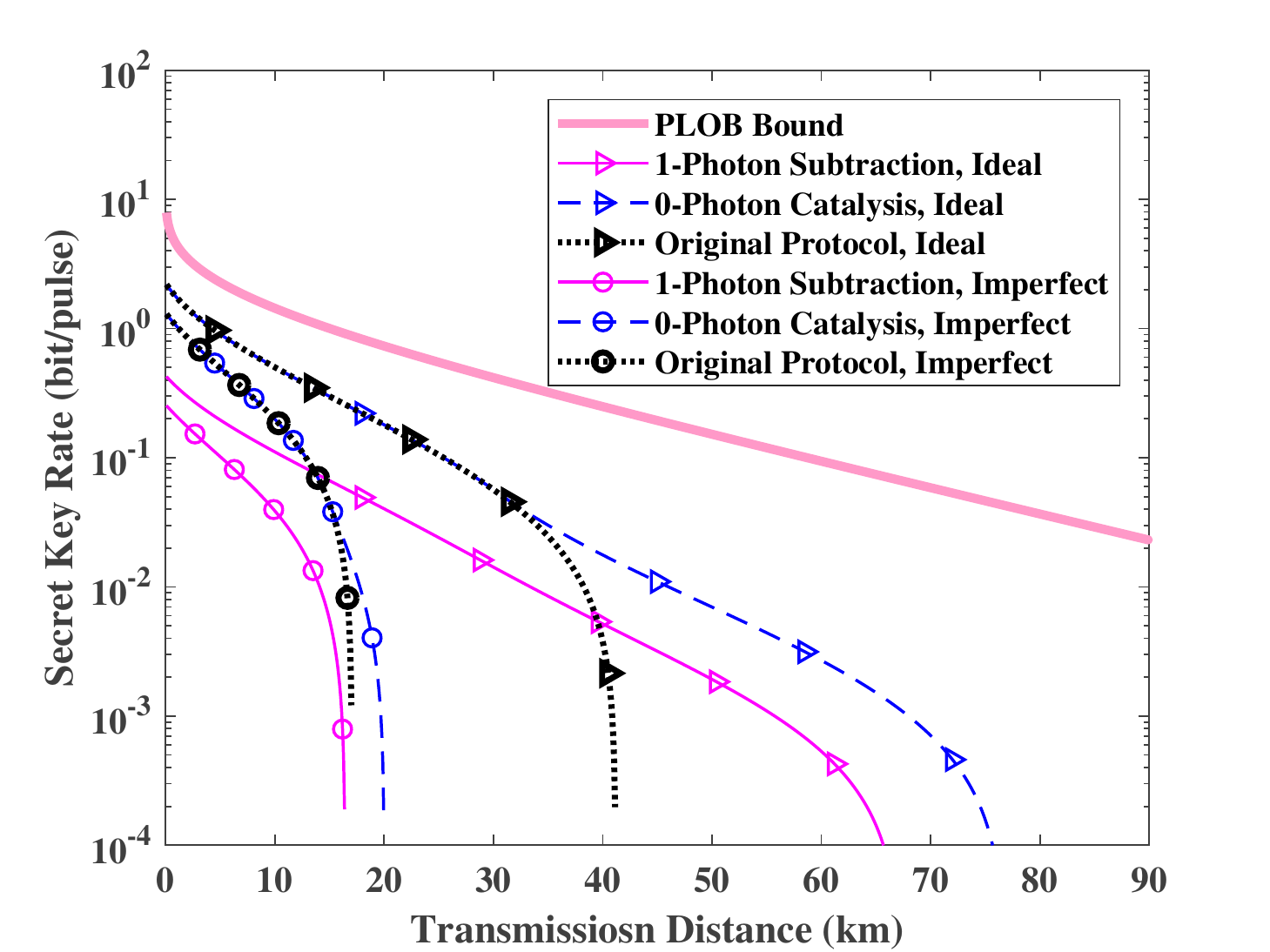}
\label{Fig4a}
}
\subfigure[]{
\includegraphics[width=\columnwidth]{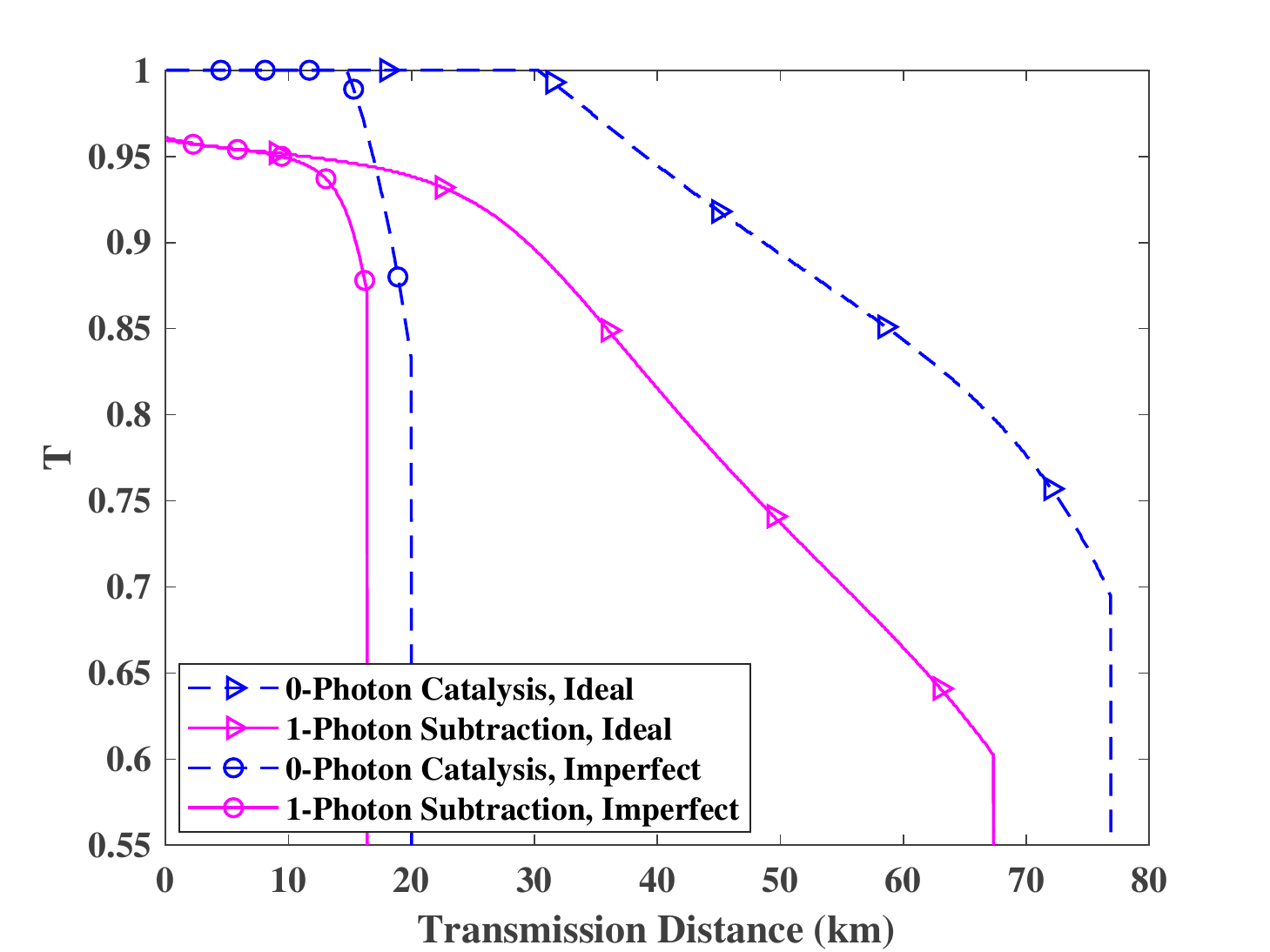}
\label{Fig4b}
}
\caption{(Color online) (a) The maximal secret key rate of the ZPC-based
CV-MDI-QKD (blue dashed line) versus the transmission distance under the
ideal and imperfect detections. (b) The optimal transmittance $T$ versus the
transmission distance corresponding to (a). To make comparisons, the black
dotted line stands for the original protocol. The thin magenta solid line
stands for the SPS-based CV-MDI-QKD. The thick pink solid line stands for
the PLOB bound.}
\label{Fig4}
\end{figure}

In the traditional CV-MDI-QKD protocols, the performance of the symmetric
case $\left( L_{AC}=L_{BC}\right) $ is worse than that of the asymmetric
case $\left( L_{AC}\neq L_{BC}\right) $ with respect to the maximal
transmission distance. In particular, for the extreme asymmetric case $%
L_{BC}=0$, the total transmission distance $L_{AB}=L_{AC}+L_{BC}$ can reach
the longest ultimate transmission distance. Based on this circumstance, we
consider the performance of CV-MDI-QKD protocol in the extreme asymmetric
case involving ZPC and SPS. To compare with the previous work \cite{32}, we
set $V_{A}=V_{B}=40,\varepsilon _{A}=\varepsilon _{B}=0.002$ \cite{13} and $%
\beta =0.95$. Moreover, taking Charlie's homodyne detection imperfections
into account, in the following we consider the ideal detection ($\eta
=1,v_{el}=0$) and the imperfect detection ($\eta =0.975,v_{el}=0.002$),
respectively.

For the optimal transmittance $T$ corresponding to Fig. 4(b), Fig. 4(a)
shows the secret key rate as a function of $L_{AB}$ involving the ideal and
imperfect detections where the black dotted line stands for the traditional
protocol. We find that the ZPC-based CV-MDI-QKD protocol (blue dashed line),
even in the imperfect detection case, can be superior to the traditional
protocol. Namely, the ZPC can be used not only for increasing the secret key
rate, but also for lengthening the transmission distance. Compared with the
performance of the original protocol, the maximal transmission distance $%
L_{AB}$ of the ZPC-based CV-MDI-QKD for a given secret key rate $10^{-4}$
bit/pulse can be extended approximately $35.6$ km for the ideal detection ($%
2.95$ km for the imperfect detection). Even if the transmittance $T$ of BS
is optimized, the performance of the proposed protocol is equivalent to that
of the original protocol at the short-transmission distance. The reason is
that there is no quantum catalytic effect when $T=1$ of the BS.

\begin{figure}[tbp]
\label{Fig5} \centering \includegraphics[width=\columnwidth]{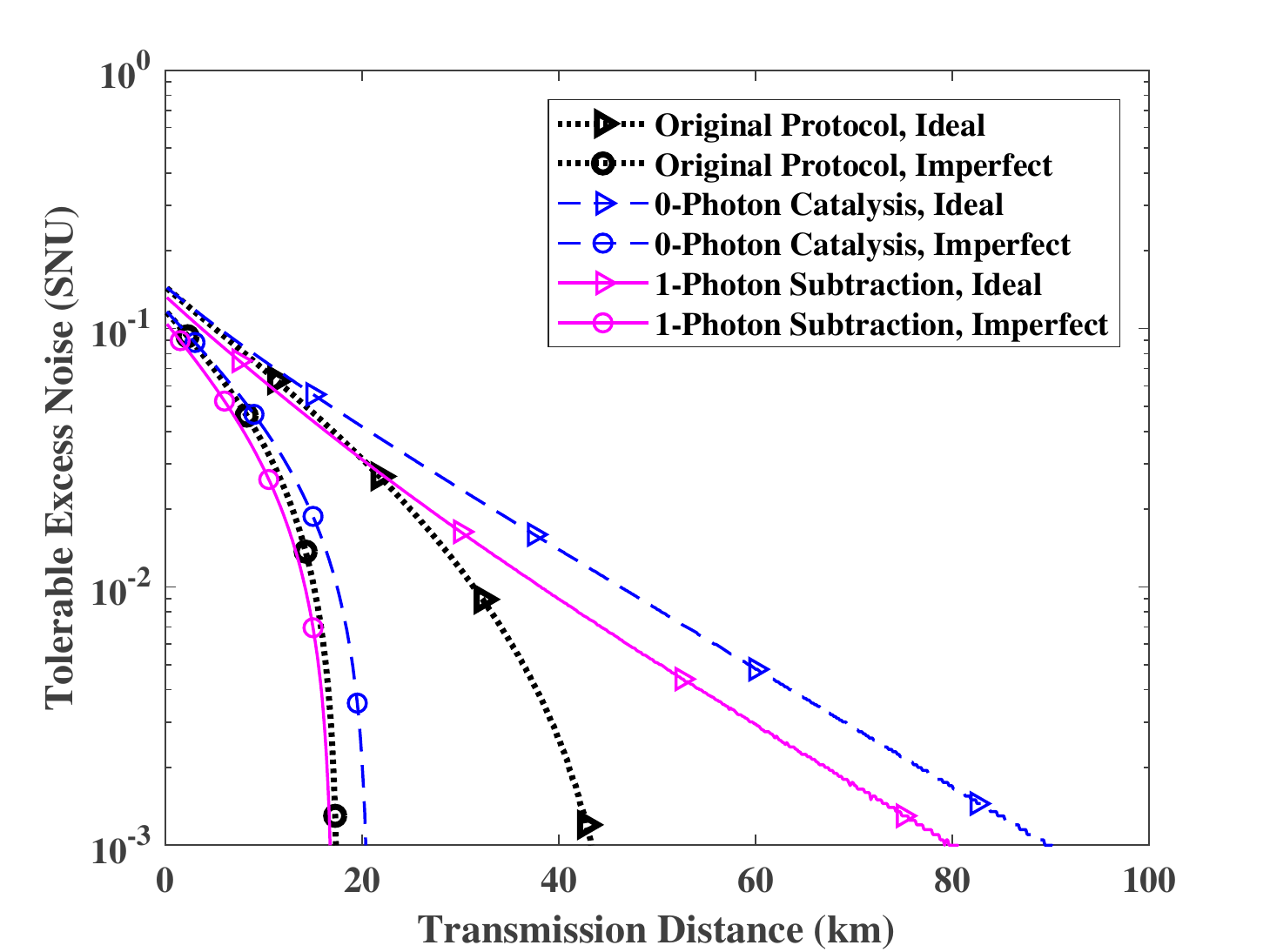}
\caption{(Color online) The maximal tolerable excess noise of the ZPC-based
CV-MDI-QKD (blue dashed line) versus the transmission distance under the
ideal and imperfect detections. As comparisons, the black dotted line stands
for the original protocol. The thin magenta solid line stands for the
SPS-based CV-MDI-QKD. }
\end{figure}
\begin{figure}[tbp]
\label{Fig6} \centering \includegraphics[width=\columnwidth]{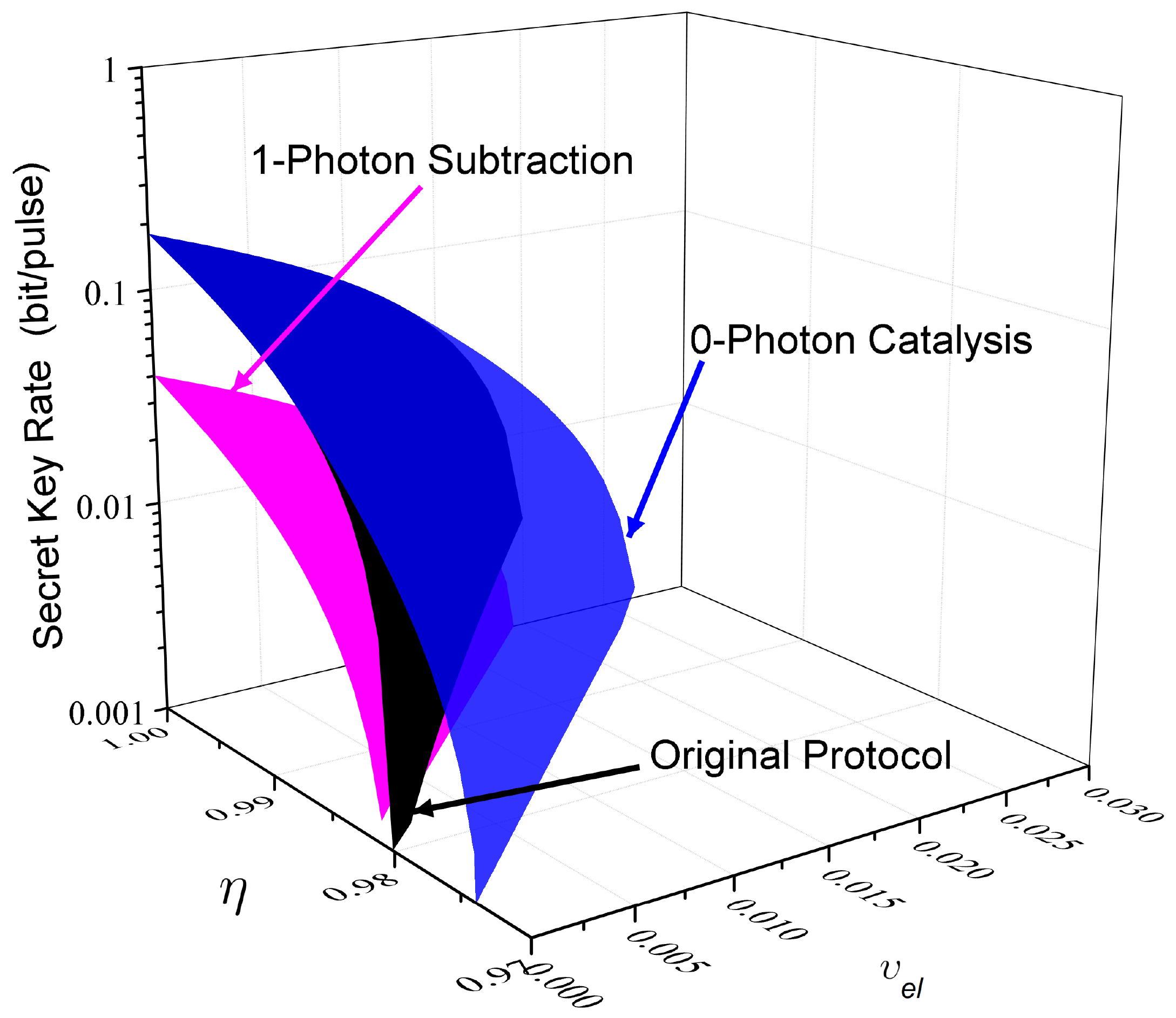}
\caption{(Color online) The secret key rate of the ZPC-based CV-MDI-QKD
protocol (blue surface), the SPS-based CV-MDI-QKD protocol (magenta surface)
and the original protocol (black surface) as a function of detection
efficiency $\protect\eta $ and electronic noise $v_{el}$ at a given
transmission distance $L_{AB}=20$ km. }
\end{figure}

Moreover, the tolerable excess noise is another common criteria for
evaluating the performance of CVQKD protocols. In Fig. 5, we illustrate the
maximal tolerable excess noise of the ZPC-based CV-MDI-QKD system as a
function of $L_{AB}$ with two cases of the ideal detection and the imperfect
detection, when optimized over the transmittance $T$. The numerical
simulation results show that, with the same parameters, the proposed
protocol presents better than other cases in terms of the maximal tolerable
excess noise. The reason is that the ZPC is indeed regarded as a noiseless
attenuation, which has been proved to increase the maximal tolerable excess
noise \cite{48}. For instance, if $\varepsilon \sim 0.001$, the proposed
protocol enables to lengthen the transmission distance up to $90$ km for the
ideal detection ($20$ km for the imperfect detection) between two remote
users, which indicates that the ZPC makes the CV-MDI-QKD protocol more
tolerant to excess noise.

In order to highlight the advantages of quantum catalysis, we show the
performance of the SPS-based CV-MDI-QKD system (thin magenta solid line)
with respect to the secret key rate and tolerable excess noise, shown in
Fig. 4(a) and Fig. 5. We find that the performance of our protocol with the
same parameters surpasses the SPS-based CV-MDI-QKD case. The reason is that
the success probability of the ZPC is higher than that of SPS, thereby
enabling to avoid the loss of information during the extraction of the
secret key rate by Alice and Bob. In addition, the physical mechanism of the
ZPC is regarded as a noiseless attenuation, which makes signal states
strongly indistinguishable to Eve and thus reduces the amount of information
stolen. Despite its appealing merits, the proposed protocol is closer to the
Pirandola-Laurenza-Ottaviani-Banchi (PLOB) bound \cite{49} that represents
the ultimate limit of repeaterless communication than the SPS-based
CV-MDI-QKD.

From the above-mentioned analysis, it shows that not only are channel
imperfections the threat to the security of CV-MDI-QKD protocols, but also
both detection efficiency $\eta $ and electronic noise $v_{el}$ can affect
the information on the secret key rate. Therefore, from this practical
viewpoint, Fig. 6 shows the performance comparisons of the ZPC-based
CV-MDI-QKD, the SPS-based CV-MDI-QKD and the original protocol as a function
of $\eta $ and $v_{el}$. It is found that under the framework of a
metropolitan area the performance of CV-MDI-QKD protocol using the ZPC is
superior to the other two cases when both detection efficiency $\eta $ and
electronic noise $v_{el}$ take a definite value. That is to say, our
protocol allows lower detection efficiency and higher electronic noise in
the case of achieving the same performance.

\section{Conclusion}

In summary, under the extreme asymmetric case, we have suggested an approach to
performance improvement of the CV-MDI-QKD involving the ZPC that can be
seen as a noiseless attenuation. We derive the secret key rate of the
the  ZPC-based CV-MDI-QKD system in the asymptotic regime. The simulation results show that
the performance of our protocol can outperform that of the original protocol
in terms of the maximal tolerable excess noise and the achievable transmission distance,
which means that exploiting such a ZPC operation can provide guidance in
designing long distance CVQKD systems. Furthermore, to highlight the
advantages of the ZPC in CV-MDI-QKD, as a comparison, the previous SPS-based
CV-MDI-QKD is presented. We find that the proposed protocol is
superior to the previous SPS case with respect to the secret key
rate, the transmission distance and the tolerable excess noise. In particular,
from a practical implementation, adding the ZPC operation makes the CV-MDI-QKD
protocol more tolerant of homodyne detection imperfections under the
framework of a metropolitan area in contrast to the SPS case. Although both
of them cannot overcome the PLOB bound at any transmission distance, this
prompts us to find other approaches (e.g., two-way CVQKD \cite{50}) to break
through the bound.

\section*{Acknowledgments}
 This work was supported by the National Natural Science Foundation of China (Grant Nos. 61572529, 61821407, 11664017), the Outstanding Young Talent Program of Jiangxi Province (20171BCB23034) and the Postgraduate Independent Exploration and Innovation Project of Central South University (Grant No: 2019zzts070).


\bigskip

\bigskip

\bigskip

\bigskip

\begin{thebibliography}{99}
\bibitem{1} V. Scarani, H. Bechmann-Pasquinucci, N. J. Cerf, M. Du\v{s}ek,
N. Lutkenhaus, and M. Peev, The security of practical quantum key
distribution, Rev. Mod. Phys. 81, 1301 (2009).

\bibitem{2} N. Gisin, G. Ribordy, W. Tittel, and H. Zbinden, Quantum
cryptography, Rev. Mod. Phys. 74, 145 (2002).

\bibitem{3} C. Weedbrook, S. Pirandola, R. Garcia-Patron, N. J. Cerf, T. C.
Ralph, J. H. Shapiro, and S. Lloyd, Gaussian quantum information, Rev. Mod.
Phys. 84, 621 (2012).

\bibitem{4} S. L. Braunstein and P. van Loock, Quantum information with
continuous variables, Rev. Mod. Phys. 77, 513 (2005).

\bibitem{5} S. Pirandola, U. L. Andersen, L. Banchi, M. Berta, D. Bunandar,
R. Colbeck, D. Englund, T. Gehring, C. Lupo, C. Ottaviani, J. Pereira, M.
Razavi, J. S. Shaari, M. Tomamichel, V. C. Usenko, G. Vallone, P. Villoresi,
and P. Wallden, Advances in quantum cryptography, arXiv:1906.01645
[quant-ph] (2019).

\bibitem{6} C. H. Bennett and G. Brassard, In proceedings of the IEEE
international conference on computers, Systems and Signal Processing,
Bangalore, India, (IEEE, New York, 1984), pp. 175--179.

\bibitem{7} M. Gessner, L. Pezze and A. Smerzi, Efficient entanglement
criteria for discrete, continuous, and hybrid variables, Phys. Rev. A 94,
020101 (2016).

\bibitem{8} V. Scarani and Renato Renner, Quantum cryptography with finite
resources: unconditional security bound for discrete-variable protocols with
one-way postprocessing, Phys. Rev. Lett. 100, 200501 (2008).

\bibitem{9} T. C. Ralph, Security of continuous-variable quantum
cryptography, Phys. Rev. A 62, 062306 (2000).

\bibitem{10} F. Grosshans and P. Grangier, Continuous variable quantum
cryptography using coherent states, Phys. Rev. Lett. 88, 057902 (2002).

\bibitem{11} J. Lodewyck, M. Bloch, R. Garcia-Patron, S. Fossier, E. Karpov,
E. Diamanti, T. Debuisschert, N. J. Cerf, R. Tualle-Brouri, S. W.
McLaughlin, and P. Grangie, Quantum key distribution over 25 km with an
all-fiber continuous-variable system, Phys. Rev. A 76, 042305 (2007).

\bibitem{12} C. Weedbrook, A. M. Lance, W. P. Bowen, T. Symul, T. C. Ralph,
and P. K. Lam, Quantum cryptography without switching, Phys. Rev. Lett. 93,
170504 (2004).

\bibitem{13} P. Jouguet, S. Kunz-Jacques, A. Leverrier, P. Grangier, and E.
Diamanti, Experimental demonstration of long-distance continuous-variable
quantum key distribution, Nat. Photonics. 7, 378--381 (2013).

\bibitem{14} F. Grosshans, Collective attacks and unconditional security in
continuous variable quantum key distribution, Phys. Rev. Lett. 94, 020504
(2005).

\bibitem{15} R. Renner and J. I. Cirac, de Finetti representation theorem
for infinite-dimensional quantum systems and applications to quantum
cryptography, Phys. Rev. Lett. 102, 110504 (2009).

\bibitem{16} H. J. Kimble, The quantum internet, Nature (London) 453, 1023
(2008).

\bibitem{17} X. C. Ma, S. H. Sun, M. S. Jiang, and L. M. Liang, Local
oscillator fluctuation opens a loophole for Eve in practical
continuous-variable quantum-key-distribution systems, Phys. Rev. A 88,
022339 (2013).

\bibitem{18} J. Z. Huang, C. Weedbrook, Z. Q. Yin, S. Wang, H. W. Li, W.
Chen, G. C. Guo, and Z. F. Han, Quantum hacking of a continuous-variable
quantum-key-distribution system using a wavelength attack, Phys. Rev. A 87,
062329 (2013).

\bibitem{19} H. Qin, R. Kumar, and R. Alleaume, Quantum hacking: Saturation
attack on practical continuous-variable quantum key distribution, Phys. Rev.
A 94, 012325 (2016).

\bibitem{20} A. Acin, N. Brunner, N. Gisin, S. Massar, S. Pironio, and V.
Scarani, Device-independent security of quantum cryptography against
collective attacks, Phys. Rev. Lett. 98, 230501 (2007).

\bibitem{21} K. Marshall and C. Weedbrook, Device-independent quantum
cryptography for continuous variables, Phys. Rev. A 90, 042311 (2014).

\bibitem{22} S. Pirandola, C. Ottaviani, G. Spedalieri C. Weedbrook, S. L.
Braunstein, S. Lloyd, T. Gehring, C. S. Jacobsen, and U. L. Andersen,
High-rate measurement-device-independent quantum cryptography, Nat. Photon.
9, 397 (2015).

\bibitem{23} X. Y. Zhang, Y. C. Zhang, Y. J. Zhao, X. Y. Wang, S. Yu, and H.
Guo, Finite-size analysis of continuous-variable
measurement-device-independent quantum key distribution, Phys. Rev. A 96,
042334 (2017).

\bibitem{24} Z. Y. Li, Y. C. Zhang, F. H. Xu, X. Peng, and H. Guo,
Continuous-variable measurement-device-independent quantum key distribution,
Phys. Rev. A 89, 052301 (2014).

\bibitem{25} X. C. Ma, S. H. Sun, M. S. Jiang, M. Gui, and L. M. Liang,
Gaussian-modulated coherent-state measurement-device-independent quantum key
distribution, Phys. Rev. A 89, 042335 (2014).

\bibitem{26} Chandan Kumar, Jaskaran Singh, Soumyakanti Bose, and Arvind,
Coherence assisted non-Gaussian measurement device independent quantum key
distribution, arXiv:1906.11799 [quant-ph] (2019).

\bibitem{27} S. L. Braunstein and S. Pirandola, Side-channel-free quantum
key distribution, Phys. Rev. Lett. 108, 130502 (2012).

\bibitem{28} H. K. Lo, M. Curty, and B. Qi, Measurement-device-independent
quantum key distribution, Phys. Rev. Lett. 108, 130503 (2012).

\bibitem{29} F. Xu, B. Qi, Z. Liao, and H. K. Lo, Long distance
measurement-device-independent quantum key distribution with entangled
photon sources, Appl. Phys. Lett. 103, 061101 (2013).

\bibitem{30} H. X. Ma, P. Huang, D. Y. Bai, T. Wang, S. Y. Wang, W. S. Bao,
and G. H. Zeng, Long-distance continuous-variable
measurement-device-independent quantum key distribution with discrete
modulation, Phys. Rev. A 99, 022322 (2019).

\bibitem{31} P. Wang, X. Y. Wang, and Y. M. Li, Continuous-variable
measurement-device-independent quantum key distribution using modulated
squeezed states and optical amplifiers, Phys. Rev. A 99, 042309 (2019).

\bibitem{32} Y. J. Zhao, Y. C. Zhang, B. J. Xu, S. Yu, and H. Guo,
Continuous-variable measurement-device-independent quantum key distribution
with virtual photon subtraction, Phys. Rev. A 97, 042328 (2018).

\bibitem{33} H. X. Ma, P. Huang, D. Y. Bai, S. Y. Wang, W. S. Bao, and G. H.
Zeng, Continuous-variable measurement-device-independent quantum key
distribution with photon subtraction, Phys. Rev. A 97, 042329 (2018).

\bibitem{34} T. J. Bartley, P. J. D. Crowley, A. Datta, J. Nunn, L. Zhang,
and I. Walmsley, Strategies for enhancing quantum entanglement by local
photon subtraction, Phys. Rev. A 87, 022313 (2013).

\bibitem{35} J. N. Wu, S. Y. Liu, L. Y. Hu, J. H. Huang, Z. L. Duan, and Y.
H. Ji, Improving entanglement of even entangled coherent states by a
coherent superposition of photon subtraction and addition, J. Opt. Soc. Am.
B 32, 2299 (2015).

\bibitem{36} Z. Y. Li, Y. C. Zhang, X. Y. Wang, B. J. Xu, X. Peng, and H.
Guo, Non-Gaussian postselection and virtual photon subtraction in
continuous-variable quantum key distribution, Phys. Rev. A 93, 012310 (2016).

\bibitem{37} A. I. Lvovsky and J. Mlynek, Quantum-optical catalysis:
generating nonclassical states of light by means of linear optics, Phys.
Rev. Lett. 88, 250401 (2002).

\bibitem{38} Y. Guo, W. Ye, H. Zhong, and Q. Liao, Continuous-variable
quantum key distribution with non-Gaussian quantum catalysis, Phys. Rev. A
99, 032327 (2019).

\bibitem{39} W. Ye, H. Zhong, Q. Liao, D. Huang, L. Y. Hu, and Y. Guo,
Improvement of self-referenced continuous-variable quantum key distribution
with quantum photon catalysis, Opt. Express 27, 17186-17198 (2019).

\bibitem{40} S. L. Zhang and X. D. Zhang, Photon catalysis acting as
noiseless linear amplification and its application in coherence enhancement,
Phys. Rev. A 97, 043830 (2018).

\bibitem{41} L. Y. Hu, J. N. Wu, Z. Y. Liao, and M. S. Zubairy, Multiphoton
catalysis with coherent state input: Nonclassicality and decoherence, J.
Phys. B: At. Mol. Phys. 49, 175504 (2016).

\bibitem{42} L. Y. Hu, Z. Y. Liao, and M. S. Zubairy, Continuous-variable
entanglement via multiphoton catalysis, Phys. Rev. A 95, 012310 (2017).

\bibitem{43} W. D. Zhou, W. Ye, C. J. Liu, L. Y. Hu, and S. Q. Liu,
Entanglement improvement of entangled coherent state via multiphoton
catalysis, Laser Phys. Lett. 15, 065203 (2018).

\bibitem{44} S. Pirandola, Entanglement reactivation in separable
environments,\ New J. Phys. 15, 113046 (2013).

\bibitem{45} M. Navascues, F. Grosshans, and A. Acin, Optimality of gaussian
attacks in continuous-variable quantum cryptography, Phys. Rev. Lett. 97,
190502 (2006).

\bibitem{46} R. Garcia-Patron and N. J. Cerf, Unconditional optimality of
gaussian attacks against continuous-variable quantum key distribution, Phys.
Rev. Lett. 97, 190503 (2006).

\bibitem{47} M. M. Wolf, G. Giedke, and J. I. Cirac, Extremality of gaussian
quantum states, Phys. Rev. Lett. 96, 080502 (2006).

\bibitem{48} J. Fiurasek and N. J. Cerf, Gaussian postselection and virtual
noiseless amplification in continuous-variable quantum key distribution,
Phys. Rev. A 86, 060302(R) (2012).

\bibitem{49} S. Pirandola, R. Laurenza, C. Ottaviani, and L. Banchi,
Fundamental limits of repeaterless quantum communications, Nat. Commun. 8,
15043 (2017).

\bibitem{50} S. Pirandola, S. Mancini, S. Lloyd, and S. L. Braunstein,
Continuous-variable quantum cryptography using two-way quantum
communication, Nat. Phys. 4, 726--730 (2008).
\end{thebibliography}
\end{document}